\documentclass[aps,prl,preprintnumbers,epsf]{revtex4}
\newcommand{\gtap}{\;{\raise.3ex\hbox{$>$\kern-.75em\lower1ex\hbox{$\sim$}}}\;}
\newcommand{\ltap}{\;{\raise.3ex\hbox{$<$\kern-.75em\lower1ex\hbox{$\sim$}}}\;}
\usepackage[dvips]{graphicx}
\newcommand{\bea}{\begin{eqnarray}}
\newcommand{\eea}{\end{eqnarray}}

\begin{document}
\preprint{hep-th/0701015v2}
\title{Callan-Symanzik equations and low-energy theorems with trace anomalies}
\author{Ji-Feng Yang\footnote{Email:jfyang@phy.ecnu.edu.cn}}
\address{Department of Physics, East China Normal University, Shanghai
200062, China}
\begin{abstract}
Basing on some new and concise forms of the Callan-Symanzik
equations, the low-energy theorems involving trace anomalies \`a
la Novikov-Shifman-Vainshtein-Zakharov, first advanced and proved
in Nucl. Phys. \textbf{B165}, 67 (1980), \textbf{B191}, 301
(1981), are proved as immediate consequences. The proof is valid
in any consistent effective field theories and these low-energy
theorems are hence generalized. Some brief discussions about
related topics are given.
\end{abstract}\maketitle

\section{Introduction}
It is well known that the powerful low-energy theorems \`a la
Novikov-Shifman-Vainshtein-Zakharov (NSVZ)\cite{NSVZ1,NSVZ2} have
played very important roles in the early studies of nonperturbative
QCD and physical properties of hadrons\cite{hadrons}. These theorems
are crucial also for many recent investigations concerning the
nonperturbative components of QCD, see, e.g.,
Refs.\cite{Fujii,Kharzeev,LitLET} and the references therein.
However, although these low-energy theorems involve the scale or
trace anomalies\cite{Trace}, the connection of these theorems with
the Callan-Symanzik equations (CSE), which are known to be
comprehensive or complete accounts for the trace anomalies and
renormalization issues, is not obvious in the original proofs given
in \cite{NSVZ1,NSVZ2}. In fact, as already noted in
Ref.\cite{NSVZ2}, in the elegant proof first given in \cite{NSVZ1}
the renormalization issue was 'simply' ignored. In the relatively
more sophisticated treatment of the renormalization issue in the
Appendix B of Ref.\cite{NSVZ2}, the proof of the low-energy theorems
was given in terms of an appropriate nonlocal correlation function
using a special regularization and yet its connection with CSE was
not established.

In this note, I wish to propose a simple and general proof of the
low-energy theorems given in Refs.\cite{NSVZ1,NSVZ2} and
\cite{migdal}, directly basing on the CSE in a form that is more
illuminating in comprehending the full account of trace anomalies
arising from renormalization. Specifically, I wish to demonstrate
the utility of the new versions of the Callan-Symanzik equations
grounded upon the concept of effective field
theories\cite{RGE_decpltheorm} and the {\em intrinsic} connections
between CSE and these low-energy theorems.

In Ref.\cite{RGE_decpltheorm}, a simple derivation of the
renormalization group equations (RGE) and CSE (merely basing on
the standard concept of effective field theories) was proposed,
with a new interpretation of running in terms of the decoupling
effects of the underlying structures. Such approach or
parametrization could provide us with a simple and hence
transparent comprehension of all possible ways that
regularization/renormalization can affect the field theoretical
computations. Also given there was various forms of the RGE and
CSE, suitable for different purposes. As will be clear shortly,
the low-energy theorems under consideration just come from one
version of CSE. In fact, due to the general validity of CSE, these
low-energy theorems should be valid in any consistent field
theories besides QCD. Therefore, these low-energy theorems
involving trace anomalies are generalized in this sense.

\section{renormalization of EFT and underlying structures}
To obtain the appropriate forms of CSE, we briefly recall some
reasonings and deductions of Ref.\cite{RGE_decpltheorm}. According
to the standard viewpoint, all the known quantum field theories
are only effective frameworks for dominant modes within a certain
range of scales, with the structures or modes at far separated
scales (i.e., underlying the effective ones) being 'ignored'. The
price paid for this 'ignorance' is the unphysical ultraviolet (UV)
and/or infrared (IR) singularities. Of course, given a complete
theoretical description where the underlying structures are
properly formulated, no such pathology should appear. In this
sense, we could view all the necessary regularization and
renormalization/factorization procedures as some sort of
substitutes or 'representations' of the underlying theory's
description of the effective field theory (EFT) sectors.

Since the complete theory is unavailable, our simplest speculation
of the underlying theory (UT)'s description of the effective modes
might be the addendum of the underlying parameters or constants
($[\sigma]$) to that of EFT ($[g]$) in the {\em finite or
renormalized} Green functions $\Gamma^{\cdots}([p],[g;\sigma])$, or,
equivalently, in the {\em finite or renormalized} generating
functional or path integral£º\bea Z([J]_{\texttt{\tiny
EFT}};[g;\sigma])&&\equiv \int D\mu_{\texttt{\tiny
UT}}\exp^{[i{\mathcal{S}}_{\texttt{\tiny
UT}}([g;\sigma];[J]_{\texttt{\tiny EFT}})]}=\int D\mu_{\texttt{\tiny
EFT}}\exp^{[i{\mathcal{S}}_{\texttt{\tiny
EFT}}([g];[J]_{\texttt{\tiny EFT}})+i\Delta
{\mathcal{S}}([g;\sigma];[J]_{\texttt{\tiny EFT}})]}.\eea Obviously,
the extra action $\Delta {\mathcal{S}}([g;\sigma];[J])]$ contains
all the necessary details that makes the description well defined,
in contrast to the simplified EFT framework. This very extra action,
or the very underlying structures, is responsible for all the
possible anomalies in the EFT terminology. In the following, we are
mainly concerned with scale or trace anomaly.

Since $\sum_g d_g g\partial_g$ (from now on, $d_{\{\cdots\}}$
denotes the canonical scale dimension of the corresponding
parameter or constant) induces the insertion of the canonical
trace of the energy-momentum tensor of an EFT, $-i\int
d^Dx\Theta(x)$, then it is straightforward to see that the
'canonical' piece $\sum_{\sigma} d_{\sigma} \sigma
\partial_{\sigma}$ in the underlying theory is the
only source of the trace anomalies to EFT, when it is expanded in
terms of the effective field operators, or, when the underlying
scales are taken to be infinite (large or small, the EFT limit).
Thus it is convenient to introduce the canonical trace for the
underlying theory in the following way\bea
\label{tildetheta}\tilde{\Theta}\equiv \Theta
+\Delta\Theta\Leftrightarrow i\{\sum_g d_g
g\partial_g+\sum_{\sigma} d_{\sigma} \sigma
\partial_{\sigma}\},\eea with $\Delta\Theta\Leftrightarrow
i\sum_{\sigma} d_{\sigma} \sigma\partial_{\sigma}$ denoting the
'canonical' underlying component, which shall appear as trace
anomalies in terms of EFT parameters. Conventionally, these
anomalies are attributed to the consequences of renormalization
procedures. Here, we could view the latter as an effective
'representation' (or substitute) of the true underlying
structures' contributions. We stress again that $\tilde{\Theta}$
is canonical in terms of the complete framework of the underlying
theory where $ \sum_g d_g g\partial_g + \sum_{\sigma} d_{\sigma}
\sigma \partial_{\sigma} $ is the sum of all the canonical scaling
transformations with respect to the parameters $[g;\sigma]$.

\section{New forms of CSE with
underlying structures} Now, for a general $n$-point Green function
$\Gamma^{(n)}$ (that is at least connected in the sense of Feynman
diagrams), the differential form for canonical scaling laws with
underlying structures should read
\begin{eqnarray}
\label{scaling1} \{\lambda \partial _\lambda  +  \sum_g d_g
g\partial_g + \sum_{\sigma} d_{\sigma} \sigma \partial_{\sigma}
-d_{\Gamma^{(n)}}\} \Gamma^{(n)}([\lambda p],[g;\sigma])=0.
\end{eqnarray}Since $\sum d_g g\partial_g$ inserts $-i\int
d^Dx\Theta(x)$, the alternative form of Eq.(\ref{scaling1})
reads,\begin{eqnarray} \label{CSEUT} \{\lambda
\partial _\lambda  + \sum_\sigma d_{\sigma} \sigma
\partial_{\sigma} -d_{\Gamma^{(n)}}\}
\Gamma^{(n)}([\lambda p],[g;\sigma]) =i\Gamma^{(n)}_{\Theta}
([0;\lambda p],[g;\sigma]).\end{eqnarray} This latter equation just
parallels the conventional CSE. To obtain the conventional form of
CSE, the next step is to expand $\sum_\sigma d_{\sigma} \sigma
\partial_{\sigma}$ into the sum of insertion of various EFT
operators with the associated 'anomalous' dimensions. It is easy
to see that, this expansion itself is exactly the general version
of the renormalization group\cite{RGE_decpltheorm}, which is a
'decoupling' theorem in the underlying theory's terminology.

Before elaborating on the EFT limit, we rewrite the CSE with
underlying structures (\ref{CSEUT}) in the following concise
form\bea\label{newCSE}\{\lambda
\partial _\lambda  -d_{\Gamma^{(n)}}\}
\Gamma^{(n)}([\lambda p],[g;\sigma]) =i\Gamma^{(n)}_{\tilde{\Theta}}
([0;\lambda p],[g;\sigma]),\eea with $\tilde{\Theta}$ given in
Eq.(\ref{tildetheta}). This is the basis for our rederivation of the
low-energy theorems.

In the EFT limit, the underlying parameters' contributions should
be replaced by appropriate 'agent' constants $[\bar c]$, at least
to balance the dimensions in necessary places. Then canonical
scaling $\sum_\sigma d_{\sigma} \sigma \partial_{\sigma}$ is
replaced by $\sum_{\bar c} d_{\bar c} {\bar c} \partial_{\bar c}$,
which in turn induces the insertion of the EFT operators
($[I_{O_i}]$) accompanied with appropriate 'anomalous' dimensions.
In formula, this is the 'decoupling' theorem of the underlying
structures (we use $\breve{P}_{\text{\tiny EFT}}$ to symbolize the
delicate EFT limit operation)
\begin{eqnarray}
\label{preRGE}
 &&\breve{P}_{\text{\tiny EFT}}\{\sum_\sigma
d_{\sigma}\sigma
\partial_{\sigma}[\cdots]\}=\sum_{\bar c} d_{\bar{c}}{\bar{c}}\partial
_{\bar{c}}[\cdots]=\sum_{O_i} \delta_{O_i}I_{O_i}[\cdots],
\end{eqnarray} with $I_{O_i}$ denoting the insertion of the EFT
operator $O_i$ and $\delta_{O_i}$ the corresponding 'anomalous
dimension' that must be a function of the EFT couplings $[g]$ and
the 'agent' constants $\{\bar c\}$. For further discussion of the
EFT contents of this expansion, please refer to
\cite{RGE_decpltheorm}. Consequently, the complete trace operator
$\tilde{\Theta}$ now becomes,
$\tilde{\Theta}=\Theta+\Delta\Theta\Rightarrow i\{\sum_g d_g
g\partial_g +\sum_{O_i} \delta_{O_i}I_{O_i}\}$ with $\Delta\Theta$
being now trace anomalies in terms of EFT operators, $\sum_{O_i}
\delta_{O_i}{O_i}$.

Introducing the operator $ {\hat{I}}_{i\tilde{\Theta}}\equiv
-\sum_g d_g g\partial_g-\sum_{O_i} \delta_{O_i}I_{O_i}$,
Eqs.(\ref{scaling1}) and (\ref{newCSE}) take the following concise
forms \bea &&\{\lambda
\partial _\lambda -{\hat{I}}_{i\tilde{\Theta}} -d_{\Gamma
^{(n)}}\}\Gamma ^{(n)}([\lambda
p],[g;\bar{c}])=0;\\
&&\{\lambda \partial _\lambda -d_{\Gamma ^{(n)}}\}\Gamma
^{(n)}([\lambda p],[g;\bar{c}])=i\Gamma_{\tilde{\Theta}}
^{(n)}([\lambda p],[g;\bar{c}]). \eea For the cases with composite
operators, we have \bea &&\label{CSE0}\{\lambda
\partial_\lambda-{\hat{I}}_{i\tilde{\Theta}} -d_{\Gamma}
\}\Gamma^{(n)} _{O_A,\cdots}([\lambda
p],[\lambda p_A,\cdots],[g;\bar{c}])=0;\\
\label{CSEop}&& \{\lambda \partial _\lambda
-d_{\Gamma}\}\Gamma^{(n)}_{O_A,\cdots}([\lambda p],[\lambda
p_A,\cdots],[g;\bar{c}])=i\Gamma^{(n)}
_{\tilde{\Theta},O_A,\cdots}([0;\lambda p],[\lambda
p_A,\cdots],[g;\bar{c}]).\eea At this stage, we should remind
that, for a generic EFT, there might be some trace anomalies from
the renormalization of composite operators in $\tilde{\Theta}$,
i.e., in the contents of $\sum_{O_i}
\delta_{O_i}{O_i}$\cite{RGE_decpltheorm}.

Here some remarks are in order. In the new versions of CSE, all
the renormalization/factorization procedures are understood to be
accomplished in the underlying theory's point of view.
Alternatively, one could view the presence of the underlying
parameters $[\sigma]$ or their 'agents' $[\bar c]$ as certain
prescription of a consistent regularization/renormalization to be
specified, provided the EFT could be consistently renormalized.
Thus, a renormalization prescription only affects EFT through the
trace anomalies and the presence of the 'agents'. All the objects
to be discussed below are understood to be already renormalized or
rendered well defined in the sense of underlying theory. For CSE,
all the nontrivial effects from renormalization are accommodated
in the trace anomalies, i.e., effected through the operation
${\hat{I}}_{i\tilde{\Theta}}$. We should also remind that the
foregoing derivation is not aiming at new results, but simply to
demonstrate that the conventional RGE and CSE could allow for a
very simple and natural interpretation from the viewpoint of the
complete theory that is well defined with the structures
underlying the effective theories.

For the purpose below, one could also resort to the conventional CSE
(in any specific scheme) and turn it into the form of
Eq.(\ref{CSEop}), bypassing the underlying theory viewpoint
advocated above. For example, in QED, this is to replace the
conventional CSE\bea &&\{\lambda
\partial _\lambda -\beta_\alpha\partial_\alpha+m(1+\gamma_m)\partial_m +
n_A\gamma_A+n_\psi\gamma_\psi-d_{\Gamma ^{(n)}}\}\Gamma
^{(n)}([\lambda p],[\alpha, m; \mu])=0, \nonumber\eea with the
operator insertion version \bea&&\{\lambda
\partial _\lambda -{\hat{I}}_{i\tilde{\Theta}}
-d_{\Gamma ^{(n)}}\}\Gamma ^{(n)}([\lambda p],[\alpha, m;
\mu])=0,\nonumber \eea or equivalently,\bea \label{QED}&&\{\lambda
\partial _\lambda -d_{\Gamma ^{(n)}}\}\Gamma ^{(n)}([\lambda
p],[\alpha, m; \mu])=i\Gamma_{\tilde{\Theta}} ^{(n)}([0;\lambda
p],[\alpha, m; \mu]).\eea  Here,
$\tilde{\Theta}=\frac{\beta_\alpha}{4\alpha}F^2 +(1+\gamma_m)m{\bar
\psi}\psi -2\gamma_\psi {\bar
\psi}i\slash\!\!\!\partial\psi=\frac{\beta_\alpha}{4\alpha}F^2
+(1+{\bar\gamma}_m)m{\bar \psi}\psi$ due to equations of motion.
(Note that, in QED, $\frac{\beta_\alpha}{\alpha}=2\gamma_A$.) Such
concise form is more useful for our purpose. What we did in the
forgoing paragraphs is just the reformulation of such conventional
CSE's in any consistent EFT in terms of the underlying theory's
perspective. The conclusions obtained below remain the same if one
adopts the conventional CSE's after turning them into the form like
Eq.(\ref{CSEop}) or (\ref{QED}).

\section{Low-energy theorems with trace anomalies}
Now, we are ready to rederive or prove the low-energy theorems
first advanced and proved in Refs.\cite{NSVZ1,NSVZ2}, using the
new versions of CSE given above. Again, all the objects involved
below are understood as being at least connected in the sense of
Feynman diagrams.

First we consider a general EFT operator $O$. Since $\Gamma_O\equiv
\langle\text{vac}|O(0)|\text{vac}\rangle$ is independent of
momentum, then the Eq.(\ref{CSEop}) for this object reduces to the
following simple one\bea
\label{LET00}-d_{\Gamma_O}\Gamma_{O}([g;\bar{c}])=i\Gamma
_{\tilde{\Theta},O}([g;\bar{c}]),\eea as $\lambda\partial_\lambda
\Gamma_O=0$. That is, ($d_O=d_{\Gamma_O}$), \bea
\label{LET01}-d_{O}\langle\text{vac}|O(0)|\text{vac}\rangle=
{\hat{I}}_{i\tilde{\Theta}}
\langle\text{vac}|O(0)|\text{vac}\rangle=i\int d^Dx
\langle\text{vac}|T\{\tilde{\Theta}(x)O(0)\}|\text{vac}\rangle.\eea
Assuming that the vacuum state is translationally invariant, then we
arrive at the following familiar form\cite{NSVZ2},\bea
\label{LET02}i\int d^Dx
\langle\text{vac}|T\{O(x)\tilde{\Theta}(0)\}|\text{vac}\rangle=
-d_{O}\langle\text{vac}|O(0)|\text{vac}\rangle.\eea Noting that the
left hand side of Eq.(\ref{LET02}) is the low-energy limit of the
correlation function $\Pi_{\tilde{\Theta}O}(Q)\equiv i\int d^Dx
e^{iqx}
\langle\text{vac}|T\{O(x)\tilde{\Theta}(0)\}|\text{vac}\rangle$,
thus the low-energy theorem\bea \label{LETNSVZ}
\Pi_{\tilde{\Theta}O}(0)=
-d_{O}\langle\text{vac}|O(0)|\text{vac}\rangle\eea follows
immediately. It is just an implicit form of the CSE for the vertex
function $\Gamma_O\equiv \langle\text{vac}|O(0)|\text{vac}\rangle$.
Here, we remind that the trace operator $\tilde{\Theta}$ contains
all the sources that break the scale invariance, both canonical
(masses) and anomalous ones. For instance, the trace operator
$\tilde{\Theta}$ of QCD with massive quarks contains canonical quark
mass operator $\sum_f m_f{\bar q}_fq_f$ besides the scale or trace
anomalies from gluon fields
($\frac{\beta(g_s)}{4g_s}G^a_{\mu\nu}G^{a\mu\nu}$) and quark fields
($\sum_f m_f{\bar\gamma}_{m_f}{\bar q}_fq_f$). In
Refs.\cite{NSVZ1,NSVZ2}, the low-energy theorem was derived in QCD
for the trace anomaly from
$\frac{\beta(g_s)}{4g_s}G^a_{\mu\nu}G^{a\mu\nu}$, the contributions
from $\sum_f m_f(1+{\bar\gamma}_{m_f}){\bar q}_fq_f$ was moved to
another side as terms that are formally linear in quark masses. That
is, denoting $\tilde{\Theta}_{\text{gluon}}\equiv
\frac{\beta(g_s)}{4g_s}G^a_{\mu\nu}G^{a\mu\nu}$ and moving $\sum_f
m_f(1+{\bar\gamma}_{m_f}){\bar q}_fq_f$ to the right hand side,
Eq.(\ref{LETNSVZ}) for QCD could be cast into the following form
\bea \Pi_{\tilde{\Theta}_{\text{gluon}}O}(0)=
-d_{O}\langle\text{vac}|O(0)|\text{vac}\rangle[1+{\mathcal{O}}(m)],
\eea which is exactly Eq.(52) in Ref.\cite{NSVZ2}.

In pure 4-dimensional gluodynamics, for
$O=\tilde{\Theta}=\frac{\beta(g_s)}{4g_s}G^a_{\mu\nu}G^{a\mu\nu}$,
the above low-energy theorem reads ($d_{\tilde{\Theta}}=4$)\bea
\label{LETNSVZQCD}
\Pi_{\tilde{\Theta}\tilde{\Theta}}(0)=-4\langle\text{vac}|
\tilde{\Theta}|\text{vac}\rangle.\eea This is the important relation
that has been extensively exploited in various QCD (including
SUSYQCD) and hadron studies\cite{Kharzeev,Fujii,LitLET}.

Applying the operation ${\hat{I}}_{i\tilde{\Theta}}$ $n$ times, we
could obtain identities\cite{NSVZ2,migdal} \bea
({\hat{I}}_{i\tilde{\Theta}})^n
\langle\text{vac}|O(0)|\text{vac}\rangle=i^n\int \prod_{i=1}^n
d^Dx_i\langle\text{vac}|T\{\prod_{i=1}^n(\tilde{\Theta}(x_i))O(0)\}
|\text{vac}\rangle=
(-d_O)^n\langle\text{vac}|O(0)|\text{vac}\rangle.\eea These
relations could be seen as corollaries of CSE.

Next, we would like to show that the low-energy theorems for the
amplitudes with non-vanishing momentum given in Ref.\cite{migdal},
are just alternative forms of the new version of CSE,
Eq.(\ref{CSEop}). These low-energy theorems are the following
relations,\bea\label{migdal3}
&&\Pi_3=2\frac{d\Pi_2}{d\ln Q^2}-(2d_O-D)\Pi_2, \\
\label{migdal4}&&\Pi_4=4 \frac{d^2\Pi_2}{(d\ln
Q^2)^2}-4(2d_O-D)\frac{d\Pi_2}{d\ln Q^2}+(2d_O-D)^2\Pi_2,\eea and
so on, where \bea\Pi_k=i^{k-1}\int d^Dx_1\cdots d^Dx_{k-1}
e^{iqx_1}\langle \text{vac}|T\{O(x_1)O(0)\tilde{\Theta}(x_2)\cdots
\tilde{\Theta}(x_{k-1}) \}|\text{vac}\rangle,\ k\geq 2, \eea and
$Q^2=-q^2$.

Again, we note that in the following and also in the foregoing
derivations, the contents of the trace operator
$\tilde{\Theta}(x_i)$ will not be specified, hence it is
applicable to both QCD and other EFT's.

To proceed, we first note that, using our notations, the left hand
sides of Eqs.(\ref{migdal3}), (\ref{migdal4}) are just
${\hat{I}}_{i\tilde{\Theta}}\Pi_2,
{\hat{I}}_{i\tilde{\Theta}}\Pi_3$, respectively. Next, we note
that $2\frac{d\Pi_2}{d\ln Q^2}=\lambda\partial_\lambda
\Pi_2(\lambda Q)$. Since $2d_O-D=d_{\Pi_2}$, so Eq.(\ref{migdal3})
becomes\bea
{\hat{I}}_{i\tilde{\Theta}}\Pi_2=i\Pi_{\tilde{\Theta},2}=
(\lambda\partial_\lambda -d_{\Pi_2})\Pi_2.\eea This is nothing but
the CSE (Eq.(\ref{CSEop})) for $\Pi_2$.

To turn Eq.(\ref{migdal4}) into the form of CSE, we employ
Eq.(\ref{migdal3}) to replace each $ 2\frac{d\Pi_2}{d\ln Q^2}$
term in Eq.(\ref{migdal4}) with $\Pi_3+d_{\Pi_2}\Pi_2$ repeatedly
till all such terms in Eq.(\ref{migdal4}) are replaced. Then we
end up with \bea
{\hat{I}}_{i\tilde{\Theta}}\Pi_3=2\frac{d\Pi_3}{d\ln
Q^2}-d_{\Pi_3}\Pi_3=(\lambda\partial_\lambda -d_{\Pi_3})\Pi_3,\eea
where obviously, $d_{\Pi_3}=d_{\Pi_2}=2d_O-D$. This is again a
CSE.

In general, we have the following CSE \bea
\Pi_{k+1}={\hat{I}}_{i\tilde{\Theta}}\Pi_k=(\lambda\partial_\lambda
-d_{\Pi_k})\Pi_k,\ \ d_{\Pi_k}=2d_O-D, \forall k \geq2.\eea

It is of course possible to derive more low-energy theorems
involving trace anomalies from the new form of CSE,
Eq.(\ref{CSEop}), by studying the low-energy limits of various
objects so that $\lambda\partial_\lambda$ does not contribute.

\section{discussions and summary}
In the above deductions, we have deliberately been not specific
about the concrete contents of the EFT trace operator and trace
anomalies. Therefore, our formulation and derivation are valid for
any kind of EFT, as long as it could be consistently regularized
and renormalized/factorized. One could well apply such low energy
theorems in the EFT's other than QCD. It would be especially
interesting to consider its possible implications for the
low-energy EFT's of QCD and/or the electroweak sectors of the
standard model.

In Ref.\cite{Kharzeev}, the low-energy theorem involving trace
anomaly in QCD has been exploited to predict 'soft' pomeron in an
impressive manner. This nonperturbative approach to hadronic
interactions at high energy and small momentum transfer is
directly based on trace anomaly of QCD. With the direct
connections between CSE and the low-energy theorems of QCD being
revealed, and the interesting relations between the low-energy
theorem approach and JIMWLK approach\cite{JIMWLK} (as an evolution
equation) being pointed out in Ref.\cite{Kharzeev}, we feel it an
interesting attempt to explore the possible connections or
interplay between CSE and JIMWLK. We would like to note that the
new form of CSE we used here look similar to the JIMWLK equation
in the sense that the scale 'evolution' $\lambda\partial_\lambda$
of CSE is governed by the operator ${\hat{I}}_{i\tilde{\Theta}}$,
while the rapidity evolution of JIMWLK is governed by the
corresponding operator defined in Ref.\cite{JIMWLK}.

Of course, the CSE's describe the full scaling laws for an EFT
with respect to all the active dynamical parameters, not a
specific or partial evolution along certain dynamical variable
(say, longitudinal momentum fraction $x$ or transverse momentum
squared $Q^2$). Nevertheless, it is natural to introduce the
'running' of the couplings of the EFT operators appearing in the
trace anomalies, i.e., $\delta_{O_i}O_i$, as is already explicated
in Ref.\cite{RGE_decpltheorm}. These 'running' objects obey the
corresponding evolution equations introduced using Coleman's
bacteria analogue\cite{coleman} with the 'anomalous dimensions'
(or, in a sense, the evolution kernels) given by the 'decoupling
theorem' or EFT expansion described by Eq.(\ref{preRGE}). It would
be interesting to explore if there is any hidden relation between
the full scaling laws encoded in CSE and the well-known evolution
equations in QCD like DGLAP\cite{DGLAP}. At least there is a
conceptual link: The treatment of soft and collinear singularities
would inevitably lead to the introduction of new scales (e.g.,
factorization scales) or dimensional parameters that should
somehow contribute to the trace anomalies. Since these evolution
equations prompt the definition of certain nonperturbative objects
(PDF, or matrix elements between hadronic states), examining these
equations from the perspective of CSE as the full scaling laws
should be helpful in clarifying the overall structure of QCD and
the like, especially in delineating the delicate interplay between
the perturbative and nonperturbative sectors. Investigations of
such possibilities will be pursued in the future. It is also in
conformity with the recent efforts using renormalization group
methods to resum various large logarithms in order to avoid
certain pathologies like Landau-pole singularity in the
conventional resummation approaches\cite{RGEresum}.

In summary, we presented some new forms of Callan-Symanzik
equations and showed that the important low-energy theorems
involving trace anomalies \`a la NSVZ follow as immediate
consequences of the new forms of CSE. In other words, these
theorems was proved in a simple and general manner so that they
are valid in any consistent EFT's. The possible relations between
CSE and various QCD evolution equations and other possible
applications of the CSE were briefly discussed.

\section*{Acknowledgement}
Some helpful conversations with Dr. I. Schienbein (Southern
Methodist U.) and Dr. M. Maniatis (Heidelberg U.) are happily
acknowledged. This project is supported in part by the National
Natural Science Foundation under Grant No. 10475028.

\end{document}